\documentstyle[12pt]{article}
\newcommand{\calA}{\mbox{${\cal A}$}}
\newcommand{\DerA}{\mbox{${\rm Der}{\cal A}$}}
\newcommand{\Aproj}{\mbox{$\calA_{proj}$}}
\newcommand{\Aconst}{\mbox{$\calA_{const}$}}

\newcommand{\RR}{\mbox{${{\bf R}}$}}

\newcommand{\ZA}{\mbox{${{\cal Z}(\calA )}$}}
\newcommand{\kwadrat}{\rule{2mm}{2mm}}

\newtheorem{lemma}{Lemma}[section]

\newtheorem{proposition}{Proposition}[section]

\newcommand{\calH}{\mbox{${\cal H}$}}

\newcommand{\kerG}{\mbox{${\rm ker}{\bf G}$}}
\newcommand{\Cstar}{\mbox{$C^*$}}

\newcommand{\za}{\mbox{${\cal Z}({\cal A})$}}

\begin{document}

\title{Noncommutative Regime of Fundamental Physics}
\author{Michael Heller\thanks{Correspondence address: ul.
Powsta\'nc\'ow Warszawy 13/94, 33-110 Tarn\'ow, Poland. E-mail:
mheller@wsd.tarnow.pl} \\
Vatican Observatory, V-00120 Vatican City State
\and Wies{\l}aw Sasin \and Zdzis{\l}aw Odrzyg\'o\'zd\'z \\
Institute of Mathematics, Warsaw University of
Technology \\
Plac Politechniki 1, 00-661 Warsaw, Poland}
\date{\today}
\maketitle
\begin{abstract}
We further develop a model unifying general relativity with
quantum mechanics proposed in our earlier papers (J. Math. Phys.
{\bf 38}, 5840 (1998); {\bf 41}, 5168 (2000)). The model is
based on a noncommutative algebra \calA \ defined on a groupoid
$\Gamma= E\times G$ where $E$ is the total space of a fibre
bundle over space-time and $G$ a Lie group acting on $E$. In
this paper, the algebra \calA \ is defined in such a way that
the model works also if $G$  is a noncompact group. Differential
algebra based on derivations of this algebra is elaborated which
allows us to construct a ``noncommutative general relativity''.
The left regular representation of the algebra \calA \ in a
Hilbert space leads to the quantum sector of our model. Its
position and momentum representations are discussed in some
detail. It is shown that the model has correct correspondence
with the standard theories: with general relativity, by
restricting the algebra \calA \ to a subset of its center; with
quantum mechanics, by changing from the groupoid $\Gamma $ to
its algebroid; with classical mechanics, by changing from the
groupoid $\Gamma $ to its tangent groupoid. We also construct a
noncommutative Fock space based on the proposed model.
\end{abstract} 

\section{Introduction}
Since the seminal work by Koszul \cite{ref1} it is known that
the standard differential geometry (on a manifold) can be
formulated in terms of a commutative associative algebra $C$,
$C$-modules and connections on these modules. $C = C^{\infty
}(M)$ is here the algebra of smooth functions on a manifold $M$,
and the $C$-module is a module of smooth cross sections of a
smooth vector bundle over $M$. The main idea of noncommutative
geometry is to follow the above formulation of differential
geometry as closely as possible with the algebra $C = C^{\infty
}(M)$ replaced by any associative, not necessarily commutative,
algebra.\cite{ref2}

General relativity is a geometric theory, and --- as it has been
noticed by Geroch \cite{ref3} --- it also can be algebraically
formulated according to the Koszul program. It seems quite
natural to try, starting from this formulation, to create a
noncommutative version of general relativity in the view of its
later unification with quantum physics. The key point is the
problem of metric. In general relativity the metric is a
dynamical variable, and the components of the metric tensor are
interpreted as gravitational potentials. The problem is that in
noncommutative geometry, in general, there is no natural way of
defining metric but in some cases the metric is unique and,
consequently, in these cases it cannot be a dynamical variable.
For instance, Madore and Mourad \cite{ref4} have proved that
this is true for a broad class of derivation based
noncommutative differential calculi. To deal with this problem a
few strategies have been elaborated.

The first of them is based on Connes' spectral calculus
\cite{ref5}. Let $M$ be a smooth compact $n$-dimensional
manifold, and let us consider a pair $(\calA, D)$ where
$\calA=C^{\infty }(M)$ and $D$ is just a ``symbol'' (for the
time being). Let further $(\calA_{\pi }, D_{\pi })$ be a unitary
representation of the pair $(\calA,D)$ in a Hilbert space
$\calH_{\pi }$ such that the triple $(\calA_{\pi }, D_{\pi },
\calH_{\pi })$ is a spectral triple (in the Connes sense). 
In such a case, there exists a unique (modulo unitary
equivalence of representations $\pi $) Riemann metric $g_{\pi }$
on $M$ such that the geodesic distance between any two points
$p, \, q \in M$ is given by
$$
d(p,q) = {\rm sup}_{a\in\calA } \{|a(p)-a(q)|: \parallel[D_{\pi
}, \pi(a)]\parallel_{{\cal B}(\calH_{\pi })}\leq 1\} 
$$ 
where ${\cal B}(\calH_{\pi })$ denotes the set of all bounded
operators on $\calH_{\pi }$. If the action is defined by $$ G(D)
= {\rm tr}_{\omega }(D^{2-n}), $$ where ${\rm tr}_{\omega }$ is
the Dixmier trace, then the unique minimum $\pi_{\sigma }$ is
the representation of the pair $(\calA, D)$ in the Hilbert space
$\calH_{\sigma } = L^2(M, S_{\sigma })$ of square integrable
spinors with $D_{\sigma }$ as the Dirac operator of the
Levi-Civita connection \cite{ref6}.

It is often stressed by Connes that ``no information is lost in
trading the original Riemann manifold $M$ for the corresponding
spectral triple'' with the proviso that ``the usual emphasis on
the points $x \in M$'' is now replaced by the spectrum of the
Dirac operator, spec$(M,D)$ with each eigenvalue repeated as
required by its multiplicity.

Another approach is based on the definition of the Riemann
metric as an inner product on a cotangent bundle \cite{ref7}.
Let us consider a spectral triple $(\calA , \calH , D)$ and the
associated differential calculus $(\Omega_D\calA, d)$ where
$\Omega_D\calA $ is the graded algebra of Connes' forms over the
involutive algebra \calA \ (which is assumed to have unit) with
the Connes deferential $d: \Omega_D^p\calA
\rightarrow \Omega_D^{p+1}\calA $. In particular
$\Omega_D^0\calA = \calA $, and the space $\Omega_D^1\calA $ is
the analogue of the space of cross sections of the cotangent
bundle.
\par
The spectral triple $(\calA , \calH , D)$ uniquely determines
the canonical Hermitian structure $\Omega_D^1\calA \times
\Omega_D^1\calA \rightarrow \calA $ by
$$
\langle \alpha, \beta \rangle_D := P_0(\alpha^* \beta)
\in \calA
$$
for $\alpha , \beta \in \Omega_D^1\calA $, where $P_0$ is the
orthogonal projector onto \calA \ as determined by the inner
product on (the completion of) $\Omega_D\calA $ defined by
$(\alpha, \beta)_0 := {\rm tr}_{\omega } (\alpha^* \beta |D|^n)$
where ${\rm tr}_\omega $ is the Dixmier trace.  The above
Hermitian structure naturally extends to $$
\langle . , .\rangle_D : \Omega_D^p\calA \times \Omega_D^p\calA
\rightarrow \calA .
$$
This Hermitian structure is weakly nondegenerate, i. e.,
$\langle \alpha , \beta \rangle _D = 0$, for all $\alpha \in
\Omega_D^1\calA $ implies $\beta = 0$, and we also assume
that if $(\Omega_D^1)'$  is a dual module, one has the
isomorphism (of right $\calA $-modules) $\Omega_D^1\calA
\rightarrow (\Omega_D^1\calA)'$ by $\alpha \mapsto \langle
\alpha, . \rangle_D$. It can be shown that such a Hermitian
inner product is in fact a {\it Riemannian metric\/} on
$\Omega_D^1$. One then defines the linear connection, develops
the corresponding differential geometry, and constructs the
Hilbert-Einstein action for ``noncommutative
gravity''\cite{ref7}.

One can also define a metric in terms of derivations of a given
algebra. Noncommutative differential algebra based on
derivations was developed by Dubois-Violette \cite{ref8}. It was
chosen by Madore to elaborate a noncommutative version of
classical gravity \cite{ref9}. If \calA \ is any associative
involutive algebra with unit, one can construct over \calA \ a
universal differential calculus $(\Omega_u\calA, d_u)$ such that
any other differential calculus over \calA \ can be obtained as
a quotient of it. Let $(\Omega
\calA, d)$  be another differential calculus over \calA . Then
there exists a unique $d_u$-homomorphism $\phi: \Omega_u \calA
\rightarrow \Omega \calA $ given by $\phi(d_uf) = df, \, f \in
\calA $, and if we know how to construct the \calA -module
$\Omega_1 \calA $ and the mapping $d: \calA \rightarrow \Omega^1
\calA$ (satisfying the Leibniz rule), then there is a method of
constructing all $\Omega^p \calA $, for $p \geq 2$, and suitably
extending the differential $d$.\cite{ref10}
\par
The idea of Madore is to define $\Omega^1 \calA $ with the help
of derivations.  He assumes that derivations are internal (which
implies that \calA \ is noncommutative). Let, for any $n \in
{\bf N}$, $\lambda_i$ be a set of $n$ linearly independent
anti-Hermitian elements of \calA . Then the derivation of
$\lambda_i $ is defined to be $e_i = {\rm ad} \lambda_i $. We
assume that if $f \in \calA $ commutes with all $\lambda_i $
then $f$ belongs to the center of \calA . The differential $d:
\calA \rightarrow \Omega_1 \calA $ is defined by
$$
df(e_i) = e_i f = [\lambda_i , f],
$$
and the \calA -bimodule $\Omega^1 \calA $ is generated by all
elements of the form $fdg$ (or $(df)g)$. We further assume that
there exist $n$ elements $\theta^i \in \Omega^1 \calA $ such
that 
$$
\theta^i(e_j) = \delta^i_j;
$$
they are called a {\it frame\/}. If a frame exists then
$\Omega^1 \calA $ is a free (left or right) module of rank $n$.
If this is the case, the construction proceeds essentially
according to the general scheme.
\par
The metric is defined by the equation
$$
g(\theta^i \otimes \theta^j) = g^{ij}
$$
where $g^{ij} \in \calA $. The metric $g$ is assumed to be
bilinear, and for $f \in \calA $ one has
$$
fg^{ij} = g(f\theta^i \otimes \theta^j) = g(\theta^i \otimes
\theta^jf) = g^{ij}f
$$
which means that the coefficients $g^{ij}$ belong to the center
\za \ of \calA . If the center \za \ is trivial, then the
coefficients $g^{ij}$ cannot be functions of coordinates, and
the metric is essentially unique. Commenting on this result,
Madore says that ``the classical gravitational field and the
noncommutative nature of space-time are two aspects of the same
thing''\cite{ref11}. Indeed, if we identify the metric with the
gravitational field and the metric is uniquely determined by the
noncommutative differential calculus, then each such
differential calculus implies the unique gravitational field.
This is true provided that the center of the algebra is trivial.
If this is not the case, the differential calculus does not
necessarily determine the metric, and the larger the center, the
larger the degree of this indeterminacy.  Therefore, in the
limit of a commutative algebra (where the center coincides with
the entire algebra) there is no obvious connection between the
differential calculus and a metric.

In a series of works \cite{ref12} we have proposed another
approach in which a noncommutative algebra is defied on a
groupoid $\Gamma = E \times G$ where $E$ is the total space of a
principal fibre bundle and $G$ is a Lie group acting on $E$. The
metric and the rest of differential geometry are based on
derivations of this algebra.  The structure of the groupoid
allows for a conceptually transparent unification of general
relativity and quantum mechanics. The point is that the algebra
on the groupoid can be naturally made a \Cstar -algebra, and
then the system can be quantized in the usual algebraic way (\`a
la Haag and Kastler).  Essentially, the $E$-part of the model is
responsible for generally relativistic effects, and the $G$-part
of the model for quantum effects. Being noncommutative our
\Cstar -algebra defines a nonlocal space. In this way, all
problems with infinities connected with sharp localization of
quantum entities are a priori avoided. The model naturally
explains effects due to correlations between distant events such
as the EPR type of experiments in quantum mechanics and the
horizon problem in cosmology. Another nice feature of the model
is that it unifies concepts which seemed to be independent of
each other. For instance, our \Cstar - algebra generates a von
Neumann algebra which defines both a generalized dynamics and a
generalized probabilistic measure (the dynamics is
probabilistic, in a generalized sense, from the very beginning).
When this dynamics is viewed from the quantum mechanical
perspective it looks like the usual Schr\"odinger unitary
evolution, but when it is viewed from the perspective of
space-time geometry it looks like a reduction of the state
vector\cite{ref13}.

In the present work we substantially develop this model but, to
make the paper self-consistent, we also briefly summarize the
previous results. In Sec. II, we define the algebra \calA \ on
the groupoid $\Gamma $ and propose the ``unitization procedure''
which ensures the existence of the correct classical limit.
Differential algebra based on derivations of the algebra \calA \
is developed in Sec. III.  This allows us to construct a
noncommutative counterpart of general relativity (Sec. IV). The
left regular representation of the algebra \calA \ in a Hilbert
space leads to the quantum sector of our model (Sec.  V.A). By
using this representation we are able to construct ``general
relativity'' in terms of operators on this Hilbert space (Sec.
V.B). In spite of the fact that our model is nonlocal (and
consequently, it has no time in the usual sense), one can define
in it a generalized dynamics in terms of derivations of the
algebra \calA \ which play the role of integral vector fields of
the considered system (Sec. V.C). Then we discuss the position
and momentum operators (Secs. V.D and V.E), and we comment on
the position and momentum representations of our model (Sec.
V.F). In section VI, we show that the model has the correct
correspondence with the standard theories, i. e., with general
relativity (by restricting the algebra \calA \ to a subset of
its center), with quantum mechanics (by changing from the
groupoid $\Gamma $ to its algebroid), and with classical
mechanics (by changing from the groupoid $\Gamma $ to its
tangent groupoid). In our view, the main drawback of the
proposed model is that, although it nicely unifies general
relativity with quantum mechanics, it is lacking the quantum
field theoretical aspect. To at least partially improve this
situation we construct, in Sec. VII, the Fock space for this
model and define noncommutative counterpart of the standard
operators on it.

\section{An Algebra on a Transformation Grou\-poid} 
Let $ E$ be the total space of a principal fibre bundle such
that the orbits of the action of the structural group $G$ on $E$
form a differential manifold $M$, interpreted as space-time. If
we want to take into account space-time singularities we must
assume that $M$ is a differential (or structured)
space.\cite{ref14} $G$ acts (to the right) on $ E$, $E\times G
\rightarrow E$.  We shall regard $\Gamma =E\times G$ as a
groupoid. The composition of elements $ \gamma_1 =(x,h)$ and
$\gamma_ 2=(xh,k)$ is defined to be $\gamma =\gamma_
2\circ\gamma_ 1=(x,hk)$, $ x\in E,\, h,k\in G$.  The source map
$ s$ and the range map $ r$ are
\[s(x,h)=(x,e), \; \; r(x,h)=(xh,e)\] 
where $ e$ is the neutral element of $G$.  We define
\[\Gamma^{(p,e)}=\{\gamma \in \Gamma:\, r(\gamma )=(p,e)\},\]
and dually
\[\Gamma _{(p,e)}=\{\gamma \in \Gamma :\, s(\gamma )=(p,e)\}.\]
In the following we shall abbreviate $\Gamma^{(p,e)}$ and
$\Gamma_{(p,e)}$ to $\Gamma^p$ and $\Gamma_p$, respectively.

\begin{lemma} $ \Gamma^p$ and $\Gamma_p$ are groups isomorphic
with $G$. $\kwadrat $ \end{lemma}
{\bf Proof.} We shall show this for $\Gamma^p$. The composition
in $ \Gamma^p$ is \[(ph^{-
1},h)\circ (pk^{-1} ,k)=(p(h k)^{-1}, hk)\in \Gamma^p,\] and
taking the inverse
\[(ph^{-1},h)^{- 1}=(p(h^{ -1})^{-1} ,h^{-1})
\in \Gamma^p;\] 
therefore, $ \Gamma^p$ is indeed a group. Let us consider the
mapping $\varphi :\Gamma \rightarrow \Gamma^p$ given by $\varphi
(h)=(ph^{ -1},h).$ The mapping $
\varphi$ is a bijection. Indeed, let $
\varphi (h_1)=\varphi (h_2)$. Hence, $ (ph_1^{- 1}$$,h_1
)=(ph_2^{ -1},h_2)$ which implies $ h_1=h_2$. The mapping $
\varphi$ is also an epimorphism. Indeed, let $
\gamma\in \Gamma^p \Rightarrow \gamma = (ph^{-1} ,h)$. Hence, $
\varphi (\gamma )=\gamma .$ $\Box$

It can be readily checked that $ \Gamma $ is a Lie groupoid. We
recall that a {\em Lie groupoid\/} (or a {\em smooth
groupoid\/}) is a groupoid $ \Gamma $ such that $\Gamma$ is a
manifold; $ \Gamma^0$ (in our case $ \Gamma^0=E$) is a Hausdorff
submanifold of $ \Gamma$; each $\Gamma^p$, $ p\in \Gamma^0$, is
Hausdorff in the relative topology; the product and inversion
maps, and the source and range maps are submersions.\cite{ref15}

Now, we define the algebra ${\cal A}_c=C^{\infty }_c(\Gamma,{\bf
C}) \cup \calA_G$, where $C^{\infty }_c(\Gamma, {\bf C})$ is the
family of compactly supported, smooth, complex valued functions
on $\Gamma $, and $\calA_G$ is the family of compactly
supported, smooth, compex valued functions lifted to $\Gamma $
from the group $G$, i.  e., $\calA_G = \pi_G^*(C^{\infty}_c(G,
{\bf C}))$ with $\pi_G:
\Gamma \rightarrow G$ the natural projection. Let $a, b \in
\calA_c $; multiplication in $\calA_c$ \ is 
defined to be the convolution
\[(a*b)(\gamma )
=\int_{\Gamma_p}a (\gamma_ 1)b(\gamma_ 1^{-1}\gamma )d\gamma_
1\] for every $\gamma\in \Gamma_p ,\,p\in E$; $d\gamma_1 $ is a
Haar measure.

If there are problems with the non-Hausdorff behavior in the
groupoid $\Gamma $ (as it could be the case when we consider
space-times with stronger types of singularities), we can take
as the algebra $C^{\infty}_c(\Gamma , {\bf C})$ the span of
complex valued functions $a$ that are smooth (or continuous)
with compact support on an open Hausdorff subset such that each
$a$ is defined to vanish outside that open Hausdorff subset. The
fact that we are considering locally compact groupoids ensures
that there is ``enough'' of such subsets (see Ref. 15, p.31-32).

In the following, the important role will be played by the
subalgebra ${\cal A}_{proj}:=pr_{M}^{*}(C^{\infty}(M))$ where
$pr_M=\pi_M\circ\pi_ E$ with $\pi_E: \Gamma \rightarrow E$ and
$\pi_M:E\rightarrow M$ being the canonical projections. As we
shall see below, this subalgebra is needed to ensure to our
model the correct transition to the classical case. ${\cal A}_{p
roj}$ is obviously isomorphic with the algebra of smooth
functions on $M$. To incorporate it into the algebra on the
groupoid $\Gamma $ we apply  the following procedure.

Having any involutive algebra \calA \ it is always possible to
construct another algebra $\calA^+ = \calA \times {\bf C}$ with
the addition $$ (a,\lambda)+(b, \mu) = (a+b, \lambda + \mu ) $$
and multiplication $$ (a,\lambda)\cdot (b, \mu) = (ab + \lambda
b + \mu a, \lambda \mu ), $$ for $a, b \in \calA ,\, \lambda ,
\mu \in {\bf C}$.  Involution is defined by $$ (a, \lambda )^* =
(a^*, \bar{\lambda }).  $$ The unit in $\calA^+ $ is ${\bf 1} =
(0,1)$, and if \calA \ has a norm, the norm in $\calA^+$ can be
defined by $$
\parallel (a, \lambda ) \parallel = {\rm max}\{\parallel a
\parallel , |\lambda |\}.
$$ It can be readily shown that \calA \ is an ideal in $\calA^+
$.  If \calA \ has no unit then $\calA^+ $ is isomorphic to the
algebra $\tilde{\calA } $, where $\tilde{\cal A }$ is the {\it
minimal unitization\/} of \calA , i. e., the minimal unital
algebra containing \calA .\cite{ref17} We shall repeat this
procedure for the algebra $\calA_c$ over each fiber of the
groupoid $\Gamma $. First, we define the algebra of complex
valued functions which are constant on the fibres $\Gamma_p$ for
every $p\in E$ $$
\calA_{const} = \{f \in C^{\infty}(\Gamma ,{\bf C}): f_p = {\rm
const},
\forall p \in E\}
$$
where $f_p$ denotes $f|\Gamma_p$, and the (bilateral) action of
$\calA_{const}$ on $\calA_c $, $\calA_c \times \calA_{const}
\rightarrow \calA $, is given by $ (a,f) \rightarrow a \cdot f,
\; \; (f,a) \rightarrow f \cdot a $ for $a\in \calA_c,\, f\in
\calA_{const}$; of course, $a\cdot f = f\cdot a$.

Now, we define the algebra $ \calA = \calA_c \times
\calA_{const} $
with the following operations
$$
(a_1,f_1)+(a_2,f_2)=(a_1+a_2, f_1+f_2),
$$ $$
(a_1,f_1) * (a_2,f_2)=
(a_1 * a_2 + f_1a_2 + f_2a_1, f_1f_2),
$$ $$
(a,f)^* = (a^*, \bar{f}).
$$
We shall also use the additive notation by writing $(a+f)$
instead of $(a,f)$. In this way, we obtain the involutive
algebra $\calA = \calA_c \times \calA_{const}$ with unit ${\bf
1}=(0+{\bf 1})$ where ${\bf 1}$ is a constant function having
the value 1.

If we restrict this procedure to a single fiber $\Gamma_p$,
$p\in E$, it is the minimal unitization of the algebra $\calA_c$
restricted to this fiber.\cite{ref17} If we perform this
procedure for the algebra on the entire groupoid $\Gamma $, it
is a non-minimal unitization of the algebra $\calA_c$ (since we
add to $\calA_c$ not only constants, but also functions constant
on fibres of $\Gamma $).

Since the algebra ${\cal A}$ plays the crucial role in our model
we shall study some of its properties. First, we shall
demonstrate that  if the group $G$ is noncompact, the functions
of \Aconst \ can be thought of as a limit of functions having
compact supports on fibres $\Gamma_p,\,p\in E$ (if $G$ is
compact, this is trivially true). To show this we first prove
the following lemma.
\begin{lemma}
On a noncompact Lie group $G$ there is a sequence of continuous
compactly supported functions $ (\phi_n)_{ n\in {\bf N}}$, {\rm
Im}$ \phi_n \in [0,1]$ such that
\begin{enumerate}
\item
${\rm s} {\rm u}{\rm p} {\rm p}\phi_ n\subset {\rm s}{\rm u}
{\rm p}{\rm p} \phi_{n+1},$
\item
$\lim_{n \rightarrow \infty}\mu ({\rm s} {\rm u}{\rm p} {\rm
p}\phi_ n)=+\infty$ where $\mu$ is the Haar measure on $G$,
\item
$\lim_{n \rightarrow \infty }\mu (\phi^{ -1}(1))= +\infty $,
\item
$\lim_{n \rightarrow \infty}b_ n=0$ where $ b_n=\mu ({\rm s}
{\rm u}{\rm p} {\rm p}\phi_ n\setminus \phi_{\mu}^{-1}(1)),$
\item
$\lim_{n\rightarrow \infty}\phi_n(g)=1$ for $ g\in G.$
$\kwadrat$
\end{enumerate}
\label{seq}
\end{lemma}
All limits are understood in the pointwise sense.  Condition (4)
says that the functions $
\phi_n$ are equal to ``almost 1'' on their supports.
\par
\noindent
{\bf Proof.} It is evident that on ${\bf R}$ there exists a
sequence $(\lambda_n)_{n\in {\bf N}}$ of smooth functions
$\lambda_n: {\bf R} \rightarrow {\bf R}$ satisfying conditions
(1) - (5). In ${\bf R}^k$ we choose the sequence $(\bar{\lambda
}_n)_{n\in {\bf N}}$ where $\bar{\lambda }_n: {\bf R}^k
\rightarrow {\bf R}$ is defined by
$\bar{\lambda}_ n(x_1,\dots ,x_k)=\lambda_ n(x_1)\cdot
\lambda_n(x_2)\cdots \lambda_n(x_k)$. The sequence $
(\bar{\lambda}_n)_{n\in {\bf N}}$ satisfies conditions (1) - (5)
of the Lemma on ${\bf R}^k$.

Let $G \subset {\bf R}^k$ be a noncompact Lie group, and let us
define the mapping $F:G \rightarrow {\bf R}^k$ by $F(g)=g$ for
$g\in G$. $ F$ is an isometric embedding (with respect to the
natural Riemannian metric). It can be readily shown that $\phi_n
=F^{*}\bar{
\lambda}_n$ satisfies conditions (1) - (5). $\Box$

\begin{proposition}
Every function of $\calA_{const}$ is a (pointwise) limit of a
sequence of functions with compact supports on every fiber
$\Gamma_p, \, p \in E$, which satisfy conditions (1) - (5) of
Lemma (\ref{seq}).  $\kwadrat$
\end{proposition} 

{\bf Proof.} Let $ f\in C^{
\infty}(
E)$. Then $ \pi_E^{*} f\in\calA_{const} $. Let further $
(\phi_n)_{ n\in {\bf N}}$ be a sequence of smooth functions on
$G$ satisfying conditions (1) - (5) of Lemma (\ref{seq}). In
such a case, the sequence $\psi_n =f\circ\pi_ E\cdot\phi_
n\circ\pi_G,$ where $\pi_E: \Gamma \rightarrow E$ is the
projection defined by $\pi_E(\gamma ) = p$ for $\gamma \in
\Gamma_p$, is the sequence of functions satisfying the
conditions of the Proposition and $\lim_{n\rightarrow
\infty}\psi_n(\gamma )=(\pi_E^{ *}f)(\gamma )\in\Aconst
$ for $\gamma \in \Gamma $. $\Box$.

We also have
\begin{lemma}
 ${\cal A}_{proj}\subset \Aconst .
\kwadrat $
\end{lemma}
{\bf Proof.} Let us assume that $ f\in {\cal A}_{proj}.$ We
define the equivalence relation ``to be in the same fiber''
$\gamma_1\sim\gamma_2\Leftrightarrow
\,\,r(\gamma_1)=r(\gamma_
2)$. If $\gamma_1\sim\gamma_2$ then $
pr_M(\gamma_1)=pr_M(\gamma_ 2)$, and consequently $f(
\gamma_1)=f(\gamma_2)$ which 
means that $f\in \Aconst $. $\Box $

The subalgebra $\calA_{proj} $ is clearly commutative; it
belongs to the center \ZA \ of the algebra \calA .

\begin{lemma}
The algebra \Aconst \ is isomorphic with the differential
structure $\calA /\Gamma $ on the space of fibres ${\cal G} =
\bigcup_{p\in E}\Gamma_p$. $\kwadrat $
\end{lemma}
{\bf Proof.} Let us define the mapping $\calA_{const} \ni f
\mapsto \bar{f} \in \calA/\Gamma $ by $\bar{f}([\gamma ]) =
f(\gamma )$. One evidently has $\bar{f}
\circ \pi_{G} = f$ where $\pi_{G}: \Gamma \rightarrow
{\cal G}$ is the projection given by $\pi_{G}(\gamma ) = [\gamma
]$.

On the other hand, we have the inverse mapping $\calA /G \ni
\bar{f} \mapsto \bar{f} \circ \pi_G = f \in \Aconst ,$ i.e., $f
\mapsto \bar{f}$ gives us the isomorphism of the algebras
\Aconst \ and $\calA /G $. $\Box $

\section{Differential Geometry of the Groupoid}
\subsection{Metric Structure}
In this subsection we consider the ${\cal Z} ({\cal
A})$-submodule 
of derivations ${\rm D} {\rm e}{\rm r}{\cal A}$ of the algebra
${\cal A}={\cal A}_{c}\times {\cal A}_{const}$ of the form
$$
V = {\rm Der}\calA_c \oplus {\rm Der}\calA_{const} \equiv V_1
\oplus V_2
$$
and we define
$$
(u \oplus v)(a,f) = (u(a),v(f)).
$$

The pair $({\cal A},V)$  is called  {\em differential
algebra\/}; it will constitute the basis for the noncommutative
geometry of the groupoid $ \Gamma $ which we develop in the
subsequent sections. At the present stage of this model
development, the choice of the differential geometry $(\calA ,
V)$ is the matter of convention, and it has been made mainly for
the sake of simplicity. Another ``natural'' choice would be to
consider the \calA -module of all derivations of the algebra
\calA . This would give us much richer geometry (but also, in
various places, open for non-unique generalizations); we hope to
deal with this case in the future.

By the {\em metric in the \ZA-module} $V$ we understand a \ZA
-bilinear mapping $g:V\times V\rightarrow \ZA $. If $ V$ is a
free module, a signature (also a Lorentz signature) can be
assigned to $g$ (it can be shown that one can also consistently
consider the case when $ g$ is degenerate\cite{ref18}).

In our case, the structure of the \ZA-module $ V$ implies the
following form  of the metric
\begin{equation}
g(u_1+u_2,v_1+v_2)=\stackrel 1g(u_1,v_1)+\stackrel {1,2}g(u_1,v_
2)+\stackrel {2,1}g(u_2,v_1)+\stackrel 2g(u_2,v_2)
\label{metric}
\end{equation}
where $u_1,v_1\in V_1$ and $u_2,v_ 2\in V_2$.

Let us consider a 1-form $$
\theta: {\rm Der}\calA \rightarrow {\cal Z}(\calA_c) \otimes
\Aconst 
$$ (we remember that ${\cal Z}(\Aconst ) = \Aconst ))$; we have
$
\theta(u_1 + u_2) = \theta(\bar{u}_1) + \theta(\bar{u}_2)
$ where bar denotes the suitable lifting. With natural
definitions $\theta_1(u_1 + u_2) := \theta (\bar{u}_1), \;
\theta_2(u_1 + u_2) := \theta (\bar{u}_2)$, 
where $u_1 \in V_{1},\, u_2 \in V_2$, we have $ \theta =
\theta_1 + \theta_2$.  Therefore, we have proved the following
Lemma.
\begin{lemma}
$A^1(\calA_c \times \Aconst ) = A^1(\calA_c ) \times A^1(\Aconst
)$ where $A^1$ denotes the family of 1-forms on a given algebra.
$\kwadrat $
\end{lemma}
In other words, for 1-forms the ``mixed components'' do not
appear.  Since $g(u, \cdot )$ is a 1-form, the above lemma
implies that the ''mixed terms'' in metric (\ref{metric})
vanish, i. e., $\stackrel{1,2}g = \stackrel{2,1}g = 0$.

If the submodules $V_{1}$ and $V_2$ are free, the above metric
tensor can be given in terms of bases.  In our case, the \ZA
-module $V_2 = {\rm Der}\Aconst $ is always free.

The algebra \calA \ is a Cartesian product of the commutative
algebra \Aconst \ and the noncommutative algebra $\calA_c$. The
commutative part leads to many metrics which should satisfy the
part of Einstein's equation coming from
\Aconst ; the noncommutative part determines the unique metric (see
Ref. 4), and the part of Einstein's equation that corresponds to
it should be solved with respect to derivations of the algebra
$\calA_c$ (as we shall see below).

Let $V^* = {\rm Hom_{\ZA }(V, \calA)}$. It is the \ZA -module of
all \ZA -valued forms on $V$. We shall assume that $V$ is
reflexive, i. e., $V= V^{**}$. Let $g$ be a metric in the \ZA
-module $V$. We define the mapping $V \rightarrow V^* $ by $$
\Phi_g(u)(v)=g(u,v)=\stackrel 1g(u_1,v_1) + \stackrel
2g(u_2,v_2).  $$ We can also write $\Phi_g =(\Phi_{\stackrel
1g}, \Phi_{\stackrel 2g})$.  We shall assume that there is the
inverse mapping $ \Phi_g^{-1}: V^* \rightarrow V $ defined by $$
\Phi_g^{-1}(\theta ) = \Phi_{\stackrel 1g}^{-1}{\theta_1} +
\Phi_{\stackrel 2g}^{-1}{\theta_2}.
$$ Since $g$ is assumed to be nondegenerate, the mapping $\Phi_g
$ is a monomorphism. In the following, the mappings $\Phi_g$ and
$\Phi_g^{-1}$ play the role analogous to that of lowering and
raising indices in the usual tensorial calculus. The set $V^+:=
{\rm Im}\Phi_g $ is the set of {\it invertible forms\/}, i. e.,
the set of forms such that $\Phi_g^{-1}(V^+)=V$.

\subsection{Connection and Curvature}
Now, we develop the noncommutative differential geometry of the
groupoid $\Gamma $. We begin with connection. For $u = u_1 +
u_2$ and $v = v_1 + v_2$, we evidently have $ [u,v] = [u_1, v_1]
+ [u_2, v_2]$. We can define the preconnection
$\nabla^{*}:V\times V\rightarrow V^{*}$, by using the Koszul
formula, in the following way
\[(\nabla^{*}_uV)(x)=\frac 12[u(g(v,x))+
v(g(u,x))-x(g(u,v)+\]
\[+g(x,[u,v])+g(v,[x,u])-g(u,[v,x]).\]
Since both the metric $g$ and the commutator $ [u,v]$ ``split''
we have
\[(\nabla^{*}_uv)(x)=(\stackrel 1{\nabla}^{
*}_{u_{1}}v_1)(x_1)+(\stackrel 2{\nabla}^{
*}_{u_{2}}v_2)(x_2).\]

Now, we define the linear connection $\nabla :{\cal
M}\rightarrow V$, where ${\cal M}=\{(u,v)\in V\times V:\nabla^{
*}_uv\in V^+\}$, by
\[\nabla =\Phi^{-1}_g\circ\nabla^{*}=\Phi^{
-1}_{\stackrel 1g}\circ\stackrel 1{\nabla}^{
*}+\Phi^{-1}_{\stackrel 2g}\circ\stackrel 2{\nabla}^{*}.\] Hence
\[\nabla_uv=\stackrel 1{\nabla}_{u_1}v_1
+\stackrel 2{\nabla}_{u_2}v_2.\]

Since $\stackrel 2g (u, v) \in \ZA $, in the connection
$\stackrel 2{\nabla}_{u_2}\!\!v_2 $ only the terms with
commutators remain (in the Koszul formula).

The curvature of the linear connection $\nabla $ is an operator
$ R: V \times V \times V \rightarrow V $ defined by $$ R(u,v)x =
\nabla_u \nabla_v x - \nabla_v \nabla_u x -
\nabla_{[u,v]}x 
$$ and $$ R(u_1+u_2, v_1+v_2)(x_1+x_2) = R(u_1,v_1)x_1 +
R(u_2,v_2)x_2.  $$

The domain of $R$ is $\{(u,v,x) \in V \times V \times V: (v,x)
\in {\cal M}, (u,x) \in {\cal M}\}$.

We assume that $V$ is a free \ZA -module, and we choose a basis
and define the trace of any linear operator $T$ in the usual
way, i. e., tr$T=\sum_{i=1}^k T_i^i$. Of course, for any linear
operator $T = T_1 + T_2$ one has $ {\rm tr} T = {\rm tr}T_1 +
{\rm tr}T_2 $.

For any pair $u,v \in V$ one defines the family of operators $
R(u,v): V \rightarrow V $ by $$ R_{u,v}(x) = R(x,u)v, $$ and the
Ricci curvature $ {\bf ric}: V \times V \rightarrow \ZA $ by $$
{\bf ric}(u,v) = {\rm tr}R_{uv}.$$ Finally, by putting $$ {\bf
ric}(u,v) = g(\RR(u),v) $$ one obtains the Ricci
operator\cite{ref19} $ \RR: V \rightarrow V $ which also
``splits'', i. e., $ \RR = \RR_1 + \RR_2$.

\section{Noncommutative General Relativity}
We define the Einstein operator $$ {\bf G} \equiv \RR + 2\Lambda
{\bf I}: V \rightarrow V $$ and assume that the generalized
Einstein equation has the form
\begin{equation} 
{\bf G} = {\bf G}_1 + {\bf G}_2 = 0.
\label{Einstein}
\end{equation}

The purely geometrical nature of the usual Einstein equation is
``contaminated'' by the non-geometric character of the
energy-momentum tensor. Since we believe in ``monistically
geometric'' (or ``pregeometric'') character of the fundamental
physical level, we assume that on this level the energy-
momentum tensor vanishes. We hope that the material component of
the cosmic stuff could be obtained from the pregeometry via some
quantum effects. However, we retain the cosmological constant
$\Lambda $ since, as some recent investigations suggest, it can
play an important role in fundamental physics. Moreover, a
simple example of our model (in which $G$ is a finite group)
shows that nonvanishing $\Lambda$ is required for the
consistency of the model (see Ref. 17).

Since the metric $g$ is defined on the \ZA -module V, the
generalized Einstein equation should, strictly speaking,
determine both $g$ and $V$. We should notice here a subtly
interplay between these two structures.  The ''horizontal part''
of the generalized Einstein's equation (${\bf G}_2 = 0$) is
essentially the ``lifting'' of the usual Einstein's equation in
space-time $M$, and all derivations $v \in V_2$ satisfy it
trivially. Therefore, to solve this equation means to find the
metric which satisfies it, just as it is the case in the
standard general relativity. On the other hand, on the submodule
$V_1$ there is essentially one metric (up to homothety) and,
consequently, to solve the ``vertical component'' of the
generalized Einstein equation (${\bf G}_1 = 0$) means to find
all derivations $v \in V_2$ which satisfy it.

The nonlocal character of the generalized Einstein equation
should be strongly emphasized. The interaction between local and
global properties of the standard general relativity were
discussed from the very beginnings of this theory. As it is well
known, one of the main motives for Einstein to create general
relativity was the idea called by him Mach's Principle. It has
many not necessarily equivalent formulations, but the underlying
idea is that local properties of space-time should be entirely
determined by its global properties. As long discussions have
finally established, this idea is only partially implemented in
the theory of general relativity\cite{ref20}. Very strong
``anti-Machian'' property in the standard geometry is the
existence of the flat (Euclidean or Minkowskian) tangent space
at any point of the considered manifold.  The tangent space, to
the great extent, determines the local properties of this
manifold independently of its global properties.  Since
noncommutative spaces are global objects, essentially having no
local properties, they could be said ``maximally Machian''
spaces.

One of the original Einstein's formulation of the Mach's
Principle postulated that the ``metric field'' should be totally
determined by some global properties (such as the mass
distribution in space-time, see Ref. 19, p. 67). In this sense
the metric $\stackrel 1g$ in our model is fully Machian since it
is uniquely (up to homothety) determined by the noncommutative
differential calculus. This is, of course, not the case as far
as the metric $\stackrel 2g$ is concerned. We believe, however,
that this is due to the simplified character of our model and
that the future noncommutative Einstein theory will be
consequently non-local (also in the ``horizontal'' part of the
model). All problems with the Mach's Principle reappear when
space-time emerges together with its rigid local properties.
This happens when the original algebra \calA \ is restricted to
the subalgebra \Aproj , as it will be discussed below.

\section{Quantum Sector of the Model}
\subsection{Left Regular Representation of the Algebra \calA }
We have now the formulation of (generalized) general relativity
in terms of the algebra \calA . Since our goal is to obtain a
unification of general relativity with quantum mechanics, we
must relate this formulation to some mathematical structures
that are employed in quantum physics. To reach this goal we
shall use the theory of the groupoid representations in a
Hilbert space. Let us then define the representation $ \pi_q:
\calA \rightarrow {\rm End }{\calH_q} $ of the algebra $\calA =
\calA_c \times \Aconst $ in the Hilbert space $\calH_q =
L^2{\Gamma_q}$ by
\begin{equation}
\pi_q(a)(\xi ) = a_q * \xi 
\label{reprpiq}
\end{equation}
where $\xi \in \calH_q$, and $a_q$ denotes $a \in \calA $
restricted to the fiber $\Gamma_q$, $p\in E$, of the groupoid
$G$; or, more precisely,
\begin{equation}
\pi_q(a+f)(\xi ) = \stackrel 1\pi_q(a)(\xi ) + \stackrel
2\pi_q(f)(\xi )
\label{repr+}
\end{equation}
where $$
\stackrel 1\pi_q(a)(\xi ) = \int_{\Gamma_q} a({\gamma_1})\xi
(\gamma_1^{-1}\gamma )d\gamma_1, $$ and $$
\stackrel 2\pi_q(f)(\xi ) = f_q \cdot \xi .
$$ We also define the ``integrated'' representation of \calA \
\begin{equation}
\pi = \bigoplus_{q\in E} \stackrel 1\pi_q + \bigoplus_{q\in
E}\stackrel 2\pi_q
\label{reprpi}
\end{equation}
in the Hilbert space ${\cal H} = \bigoplus_{q\in E}{\cal H}_q$.
Here, and above, we use the additive convention (summation is
understood as a pair). The representation $\pi$ is what in the
theory of groupoid representations is called the {\it left
regular representation\/} of the algebra \calA \ in the Hilbert
space $\calH$ (see Ref. 15, Chapter 3.1).

Our aim is now to introduce a norm in the algebra \calA . To do
so we restrict the algebra \Aconst \ to the subalgebra
$A^b_{const}$ of {\it bounded\/} functions constant on the
fibres of the groupoid $\Gamma $. As we shall see later (Sec.
VI.A), the bounded functions of \Aconst \ are enough to ensure
the correct transition from our model to the ordinary space-time
geometry.  We define the norm in the algebra \calA $$
\parallel(a, f)\parallel = {\rm max}\{\parallel a \parallel,
\parallel f \parallel \}
$$ where $\parallel a \parallel = {\rm sup}_{q\in E} \parallel
\stackrel 1\pi_q \!\!(a)\parallel $, and $\parallel f \parallel
= {\rm sup}_{q\in E} \parallel \stackrel 2\pi_q\!\! (f)
\parallel ,\, a \in
\calA_c,\, f \in \calA^b_{const}$. It is
also the norm when $\pi $ is restricted to $\pi_q$, i. e., to a
single fibre of $\Gamma $ over $q \in E$. The algebra \calA \
completed with respect to this norm is a $C^*$-algebra. In the
following, we shall always assume that this is the case.

Since now $\calA $ is a $C^*$-algebra we can use it to quantize
the system with the help ot the algebraic method (as it is done
in Ref. 12). If $a$ is a Hermitian element of $\calA $ and
$\varphi $ a state on the algebra $\calA $, then $\varphi (a)$
is the expectation value of the observable $a$ if the system is
in the state $\varphi $. However, in the following we shall
develop the quantum sector of our model in terms of operators on
a Hilbert space (by using the above left regular representation
of \calA ) rather than directly in terms of the $C^*$-algebra.

\subsection{General Relativity in Terms of Operators on a
Hilbert Space} Let us consider any representation of the algebra
$ {\cal A}$ in the Hilbert space $\bigoplus_{p\in
E}L^2(\Gamma^{p})$, and let us assume that $\pi$ is a
monomorphism. We define $\hat {{\cal A}}:=\pi ({\cal A})$, and
also $\mbox{if $v\in {\rm D}{\rm e}{\rm r} {\cal A},$}$ $\hat
{v}(\pi (a)):=\pi (v(a))$ for $a\in {\cal A}$.

Let $A$ be a tensor of the type $(n,0)$,
$A:\DerA\times\cdots\times\DerA\rightarrow {\cal A}$. We
evidently have $\hat {A}(\hat {v}_1,\ldots ,\hat {v}_n )=\pi
(A(v_1,\ldots ,v_n))$, and similarly for $A:\DerA\times ,\cdots
,\DerA\rightarrow\ZA $, $\hat {A}:{\rm D}{\rm e}{\rm r}\hat
{{\cal A}}
\times ,\cdots ,\times {\rm D}{\rm e}{\rm r}
{\cal A}\rightarrow {\cal Z}(\hat {{\cal A}} )$, we also obtain
${\cal Z}(\hat {{\cal A}})=\widehat{\ZA }$.  And analogously,
for tensors of other types. In particular, the above refers to
derivations. We have $\hat {v}(\hat {a})=\widehat{v(a)}$ for
$a\in {\cal A}$, and since in the space of operators all
derivations are internal, ${\rm D}{\rm e}{\rm r}\hat {{\cal
A}}=\{{\rm a}{\rm d}\hat {a}:a\in {\cal A}\}$.  Hence $\hat
{v}(\hat {a})=({\rm a}{\rm d}\hat { b})(\hat {a})$, or
equivalently
\begin{equation}
\pi (v(a))=[\pi (b),\pi (a)]. \label{der}
\end{equation}

In this way, we have shown that each tensor on derivations of
the algebra ${\cal A}$ uniquely determines a tensor on operators
(in the image of the representation $\pi $). Notice that this is
valid only if $\pi$ is a monomorphism; therefore, in the case
when $\pi $ is the representation of the algebra \calA \
considered in the preceding subsection, this is valid for the
``integrated'' representation $\pi =\bigoplus_{ p\in E}\pi_p$,
and is not valid for representations $\pi_p,\,p\in E$.

In the preceding section, we have defined the generalized
Einstein equation in terms of the algebra $\calA $; now, we are
able to ``transfer'' this equation, with the help of the left
regular representation, to the Hilbert space $\calH =
\bigoplus_q L^2(\Gamma^q)$, i. e., we are able to express the
generalized Einstein equation in terms of operators on the
Hilbert space \calH . For instance, the (Lorentz) metric $g(u,
v),\, u,v \in {\rm Der}\calA $ can be expressed in the following
way $$
\hat g({\hat u},{\hat v}) = \hat g(\hat u_1 + \hat u_2, \hat v_1
+ \hat v_2) = \;\stackrel 1g({\rm ad} \hat a_1, {\rm ad} \hat
b_1) + \stackrel 2g(\hat u_2, \hat v_2), $$ and similarly for
other tensors appearing in the generalized Einstein equation.
Finally, this equation will have the form
\begin{equation}
\label{HGR}
\widehat{\bf G} = 0.
\end{equation}

It is worthwhile to notice that this formulation of the
``generalized general relativity'' has a strong nonlocal flavor:
it can be only done on the ``integrated space'' $\bigoplus_q
L^2(\Gamma^q)$, but not on a single ``fibre'' $L^2(\Gamma^q)$.

\subsection{Generalized Quantum Dynamics}
Derivations of the algebra $\hat{\calA } = \pi (\calA )$ have
the form of equation (\ref{der}). If we restrict this equation
to a single fibre of the groupoid $\Gamma$ over $q\in E$, we
obtain
\begin{equation}
\pi_q(v(a))=[\hat {b},\pi_q(a)]. 
\label{derq}
\end{equation}
We can see that equation (\ref{derq}) (resp. (\ref{der})) bears
strong resemblance to the known Schr\"odinger equation in the
Heisenberg picture of quantum mechanics, describing evolution of
observables when states are constant. Indeed, equation
(\ref{derq}) (or (\ref{der}))  can also be regarded as
describing the evolution of the operator $\pi_q(a), \, a\in
\calA $ with $\hat b$ playing the role of a ``Hamiltonian''
(which in this case depends on $v$). Although in our model there
is no concept of time (in the usual sense), derivations can be
thought of as counterparts of vector fields, and equation
(\ref{derq}) (or (\ref{der})) as modelling the dynamics in terms
of the ``integral vector field'' $v\in {\rm Der}\calA $.  Basing
on this heuristic argument we postulate that the dynamics of a
quantum gravitational system is described by the following
equation
\begin{equation}
i\hbar\pi_q(v(a))=[F_v,\pi_q(a)]\label{dyneq}
\end{equation} 
for every $q\in\ E$. Here, for the sake of generality, we assume
that $F_v$ is a one-parameter family of operators $F_v\in {\rm
E}{\rm n}{\rm d} {\cal H}$, ${\cal H}=L^2(\Gamma_q)$, such that
$$ F_{\lambda_1v_1+\lambda_2v_2}=\lambda_1F_v+\lambda_ 2F_2 $$
with $\lambda_1,\,\lambda_ 2\in {\bf C}$. To connect this
dynamics with Einstein equation (\ref{Einstein}) we additionally
postulate that the derivations $v, v_1$ and $v_2$ be solutions
of the ``vertical part'' of Einstein's equation, i. e., $v, v_1,
v_2 \in {\rm ker}{\bf G}_1$. We also assume that $[F_v,
\pi_q(a)]$ is a bounded operator. The coefficient $i\hbar $ has
been added to guarantee the correspondence with the standard
quantum mechanics. Let us notice that in fact we have a ${\bf
C}$-linear mapping $ \Phi: {\rm ker}{\bf G} \rightarrow {\rm
End}{\cal H} $ satisfying
\begin{equation}i\hbar\pi_q(v(a))=[\Phi(v),\pi_q(
a)]\label{dyneq1}.\end{equation}

It should be emphasized that the above described
``noncommutative dynamics'' depends on the ``form'' $\Phi $.
This remains in consonance with the result obtained in our
previous work\cite{ref22} in which the dynamics for our model
has been introduced in terms of von Neumann algebras. These
algebras turn out to be natural ``dynamical objects'' in
noncommutative geometry (see Ref. 2, p.44). To be more precise,
if the operator $F_v$ in Eq.  (\ref{dyneq}) is positively
defined and bounded then, on the strength of the Tomita-Takesaki
theorem\cite{ref23a}, there exists a one-parameter group
$(\alpha^{\phi }_t)_{t\in {\bf R}}$ of automorphisms of the von
Neumann algebra $\pi(\calA )^{\prime \prime}$ where $\pi =
\bigoplus_{\pi \in E} \pi_q$ (depending on a
form $\phi $ on this algebra) in terms of which dynamics can be
defined (see Eq. (6) in Ref. 20).

The fact that $v\in\kerG$ makes of eqs.  (\ref{Einstein}) and
(\ref{dyneq1}) a ``noncommutative dynamical system''.  To solve
this system means to find the set
\[{\cal E}_{\bf G}=\{a\in \calA: i\hbar
\pi_q(v(a))=[\Phi(v),\pi_ q(a)],\forall v\in {\rm k}{\rm e}{\rm
r}{\bf G} \}.\] It can be easily verified that it is a
subalgebra of \calA .

Let $\bar{\cal E}_{\bf G}$ be the smallest closed involutive
subalgebra of the algebra \calA \ containing ${\cal E}_{\bf G}$.
$\bar{\cal E}_{\bf G}$ is said to be {\em generated by} ${\cal
E}_{\bf G}$.  Since \calA \ is assumed to be a
\Cstar-algebra and every closed involutive subalgebra of a 
\Cstar-algebra is a \Cstar-algebra\cite{ref24}, $\bar{\cal
E}_{\bf G}$ is also a \Cstar-algebra; it will be called {\em
Einstein \Cstar-algebra\/} or simply {\em Einstein algebra}, and
the pair $ (\bar{\cal E}_{\bf G}, {\rm ker}{\bf G})$ -- {\em
Einstein differential algebra}.  We can also define another
subalgebra of \calA \
\[{\cal E}_v:=\{a\in \calA : i\hbar
\pi_q(v(a))=[\Phi(v),\pi_q(a)]\},\]  
where $v\in {\rm ker}{\bf G}$.  It can be easily seen that
${\cal E}_{\bf G}=\bigcap_{v\in {\rm ker}{\bf G}} {\cal E}_v$.
Of course, it can happen that ${\cal E}_ v=\{0\}$ which would
imply that also ${\cal E}_{\bf G}=0$.

Dynamical equation (\ref{dyneq}) can be written in the form $$
\pi_q(v_1(a)) + (v_2(f)) = [F_{v_1+v_2}, \pi_q(a+f)]
$$ where $a \in \calA_c , \, f \in \Aconst ,\, v_1 \in V_1, \,
v_2
\in V_2$. This is equivalent to
$$
\stackrel 1\pi_q(v_1(a)) + \stackrel 2\pi_q(v_2(f)) = [F_{v_1},
\stackrel 1\pi_q(a)] + [F_{v_2}, \stackrel 2\pi_q(f)],
$$ and the last commutator vanishes since $f$ is constant on the
considered fiber. Finally, we have $$
\pi_q(v_1(a)) = [F_{v_1}, \pi_q(a)]
$$ and $$
\pi_q(v_2(f)) = 0.
$$ The last equality implies that $(v_2(f))_q = 0$. Analogous
results are valid for the ``integrated representation'' $\pi $.

\subsection{Position and Momentum Operators} 
In this subsection we consider position and momentum operators
in our ``generalized quantum mechanics''.  Let $M$ be any
relativistic space-time (4-dimensional smooth manifold). It is
evident that the projection $ pr:\Gamma \rightarrow M$,
$pr=\pi_M\circ \pi_E$, which is clearly connected with
localization in $ M$, is not a numerical function (it has no
values in {\bf R} or {\bf C}), and consequently it does not
belong to the algebra ${\cal A})$.  However, if we choose a
local coordinate map $x = (x^{\mu }),\,
\mu = 0,1,2,3$, in $ M$ then the projection $pr$ determines four
observables in the domain ${\cal D}_x$ of $x$ $$ pr_{\mu } =
x^{\mu } \circ pr $$ We thus have  the system of four position
observables  $pr=(pr_0,pr_1,pr_2,pr_3)$. Of course, $pr_{\mu}\in
{\cal A}_{ proj}(pr^{-1}({\cal D}_x))$ and it is Hermitian.

Let us notice that  the projection $ pr:\Gamma \rightarrow M$
contains, in a sense, the information about all possible local
observables $ pr_{\mu}$. This can be regarded as a
``noncommutative formulation'' of the fact that there is no
absolute position but only the position with respect to a local
coordinate system.

Let us apply representation (\ref{repr+}) to the position
observable, e. g, to $pr_1$. Of course, $pr_1 \in \Aconst
(pr^{-1}({\cal D}_x))$, and we have $$ (\pi_q(pr_1))(\xi ) =
\stackrel 1\pi_q(0)(\xi ) + \stackrel 2\pi_q(pr_1)(\xi ) =
(pr_1)_q \cdot \xi = x \cdot \xi, $$ $x \in M$, in the local
map. We see that the position observable in the quantum sector
of our model has the same form as in the ordinary quantum
mechanics. This indicates that we are working in the position
representation of our model.

By analogy with the ordinary quantum mechanics a derivation of
the algebra ${\cal A}$ should in our model play the role of the
momentum operator.  We shall see that this is indeed the case.
However, first let us prove the following Lemma.

\begin{lemma}  Let $\pi :
{\cal A}\rightarrow {\rm E}{\rm n} {\rm d}{\cal H}$ be a
(nondegenerate) representation of the algebra ${\cal A}$ in the
Hilbert space $ {\cal H}$.  Any internal derivation $ {\rm
a}{\rm d}a,\,$$a\in {\cal A}$, of $ {\cal A}$ has an operator
representation of the form $ {\rm a}{\rm d}\pi (a)$, which we
shall also denote by $\pi ({\rm a}{\rm d}a)$.  The operator $
\pi ({\rm a}{\rm d}a)$ is an element of the algebra $
\pi ({\cal A})$ if and only if 
$a\in {\rm k}{\rm e}{\rm r}\pi$.
\end{lemma}

{\bf Proof.\/}  It is trivial to check that $ {\rm a}{\rm d}\pi
(a)$ is a derivation of the algebra $
\pi ({\cal H})$ 
(and consequently of the algebra $ {\rm E}{\rm n}{\rm d}{\cal
H}$).  For any $
\omega\in {\rm E}{\rm n}{\rm d}
{\cal H}$, ${\rm a}{\rm d}\pi (a)(\omega )=\pi (a)\circ\omega
-\omega\circ
\pi (a).$ If $a\in {\rm k}\,{\rm e}
{\rm r}\pi$, then $\pi (a)=0$.  In such a case, ${\rm a}{\rm
d}\pi (a)=\pi (a) =0$.  On the other hand, if $\pi({\rm ad}a)
\in \pi (\calA )$ then $\pi({\rm ad}a) = \pi (b)$, or  ${\rm ad}\pi
(a) = \pi (b)$ which implies that $\pi(a) = 0$. $\Box$

From this lemma it follows that the derivation, which is to be
interpreted as the momentum operator, must be an element of $
{\rm E}{\rm n}{\rm d}{\cal H}\setminus
\pi ({\cal A})$; in other 
words, it must be an external derivation.  It also must be a
``lifting'' of a local basis in space-time given by a local
coordinate system.  Let $(pr_0,pr_1,pr_2,pr_3)$ be the position
observables with respect to a local coordinate system
$x=$$(x_{}^0 ,x^1,x^2,x^3)$ in space-time $ M$.  The lifts
$\bar{\partial}_ 0,\bar{\partial}_1,\bar{\partial}_
2,\bar{\partial}_3$ (to $\Gamma $) of the basis fields
$\partial_ 0,\partial_1,\partial_2,\partial_ 3$ in $M$,
corresponding to the coordinate system $x$, satisfy the
condition
\[
\bar{\partial}_{\mu}(pr_{\nu}
)=\delta_{\mu\nu},
\]
$\mu ,\,$$\nu =0,1,2,3$. Let now
$\hat{\partial}_0,\hat{\partial}_
1,\hat{\partial}_2,\hat{\partial}_ 3$ be derivations in the
space of operators ${\rm E}{\rm n}{\rm d} {\cal H}$, where
${\cal H}=L^2 (\Gamma_{{\cal D}_x})$ and ${\cal D}_x$ is the
domain of the coordinate map $ (x_0,x_1,x_2,x_3)$. We have

\begin{lemma} 
The following commutation relations are valid
\[[\hat{\partial}_{\mu},\pi (p
r_{\nu})]=\delta_{\mu\nu}{\bf 1}
.\]
\end{lemma}

{\bf Proof.\/} Direct computation by assuming that $
\xi\in L^2(G_{{\cal D}_x})$ is a function on 
$pr^{-1}({\cal D}_x)$. $\Box$

In the following Section we shall present the above results in
an elegant mathematical form.

\subsection{A Sheaf Structure on the Groupoid}
On the Cartesian product $\Gamma=E\times G$ there exists the
natural product topology; however, we shall consider a weaker
topology in which the open sets are of the form $\pi_E^{- 1}(U)$
where $U$ is open in the manifold topology $\tau_E$ on $E$.
Every such open set is also open in the topology
$\tau_E\times\tau_G$. Indeed, every such set is given by
$\pi_E^{-1}(U)=U\times G$.

Let $\underline {{\cal A}}$ be a functor which with an open set
$ U\times G$ associates the involutive noncommutative algebra
${\cal A} (U\times G)$ of smooth compactly supported complex
valued functions with the ordinary addition and convolution
multiplication.  As it can be easily seen, $\underline { {\cal
A}}$ is a sheaf of noncommutative algebras on the topological
space $(\Gamma ,\pi_E^{-1} (\tau_E))$.

The projection $pr\!\!:\Gamma \rightarrow M$ can be {\em
locally\/} interpreted as a set of (local) cross sections of the
sheaf $\underline { {\cal A}}$ (i.e.  as a set of position
observables).  Indeed, for the domain ${\cal D}_x$ of any
coordinate map $ x=(x^0,x^1,x^2,x^3)$, the composition $x\circ
pr=(x^0\circ pr,\,x^1\circ pr,\,x^2\circ pr,\,x^3\circ pr )$ is
a set of such local cross sections of $\underline {{\cal A}}$ on
the open set $\pi_E^{-1}({\cal D}_x\times G).$ The global
mapping $ pr:\Gamma \rightarrow M$ is not a cross section of
$\underline { {\cal A}}$.

Let us notice that to a measurement result which is not a number
but a set of numbers (a vector, a spinor...  ) there does not
correspond a single observable but rather a set of observables,
i.  e.,  a set of (local) cross sections of the sheaf
$\underline { {\cal A}}$.

Now, we define the {\em derivation morphism\/} of the sheaf
$\underline { {\cal A}}$ over an open set
$U\in\pi_E^{-1}(\tau_E)$ as a family of mappings $
X=(X_W)_{W\subset U}$ such that $X_W:\underline {\calA
}(W)\rightarrow\underline { {\cal A}}(W)$ is a derivation of the
algebra $\underline { {\cal A}}(W),$ and for any $W_ 1,W_2$ open
and $W_1\subset W_2\subset U$, the following diagram commutes
\vspace{1.5cm}

\begin{center}
\vspace{1.5cm}
\hspace{-3cm}
\unitlength=1.00mm
\special{em:linewidth 0.4pt}
\linethickness{0.4pt}
\begin{picture}(87,50)(35,15)
\put(58.00,82.00){\makebox(0,0)[cc]{$\underline{\cal A}(W_2)$}}
\put(123.00,82.00){\makebox(0,0)[cc]{$\underline{\cal A}(W_2)$}}
\put(123.00,47.00){\makebox(0,0)[cc]{$\underline{\cal A}(W_1)$}}
\put(58.00,47.00){\makebox(0,0)[cc]{$\underline{\cal A}(W_1)$}}
\put(64.00,82.00){\vector(1,0){53.00}}
\put(117.00,82.00){\vector(0,0){0.00}}
\put(64.00,47.00){\vector(1,0){53.00}}
\put(58.00,76.00){\vector(0,-1){23.00}}
\put(123.00,76.00){\vector(0,-1){23.00}}
\put(90.00,87.00){\makebox(0,0)[cc]{$X(W_2)$}}
\put(90.00,51.00){\makebox(0,0)[cc]{$X(W_1)$}}
\put(52.00,65.00){\makebox(0,0)[cc]{$\rho^{W_2}_{W_1}$}}
\put(129.00,65.00){\makebox(0,0)[cc]{$\rho^{W_2}_{W_1}$}}
\end{picture}
\end{center}

\vspace{-2cm}

\noindent
where $\rho^{W_1}_{W_2}$ is the known restriction homomorphism.
The family of all derivation morphisms indexed by open sets is a
sheaf of ${\cal Z}({\cal A})$-modules where $ {\cal Z}({\cal
A})$ denotes the sheaf of centers of the algebras $\underline {
{\cal A}}(U),\,$$U\in\pi_E^{-1} (\tau_e)$.

Components of the momentum observable $
\bar{\partial}_{\mu}$ are cross sections of the sheaf 
of ${\cal Z}({\cal A})$-modules of derivations of the sheaf
$\underline {{\cal A}}$ over domains of coordinate maps, and the
representation $\pi_U:\underline {{\cal A}}(U)
\rightarrow\pi_U(\underline {{\cal A}}
(U))$, where $U\in\pi_E^{-1}(\tau_E)$, transfers the sheaf
structure from the groupoid $\Gamma $ to the family of operator
algebras over the topological space $(\Gamma ,\,$$\pi_
E^{-1}(\tau_E)$).

\subsection{Momentum Representation of the Model}
As it was noticed above, so far we were working in the position
representation of our model; in the present Subsection we shall
briefly indicate how its momentum representation can be
constructed. To this end, we must turn, by close analogy with
the standard case, to the harmonic analysis on a groupoid.

The usual Fourier transform (on the real line) changes
translations into multiplications by a function and,
consequently, it enables one to perform the spectral
decomposition of any operator which commutes with translations.
The generalized Fourier transform plays the same role with
respect to groups. Let $G$ be a topological group, and $\hat{G}$
the set of equivalence classes of irreducible unitary
representations of $G$. For every $\lambda \in
\hat{G}$ let $T_{\lambda }$ denotes a representation of
$G$ in a Hilbert space $\calH_{\lambda }$ which belongs to this
equivalence class. Let us also assume that $G$ is a locally
compact group, and let us consider $f\in L^1(G, dg)$. The
operator valued function $\tilde{f}:\hat{G} \rightarrow {\rm
End}\calH_{\lambda }$ defined by $$
\tilde{f}(\lambda ) = \int_{G}f(g)T_{\lambda }(g)dg
$$ is said to be the {\it Fourier transform\/} of $f$ at
$\lambda
\in \hat{G}$.

If $G$ is a compact noncommutative group, the set $\hat{G}$ is
discrete\cite{ref25}.  Since all irreducible representations of
$G$ are finite dimensional, we can assume that $\tilde{\lambda
}\in {\rm Mat}_{n(\lambda )}({\bf C})$ where $n(\lambda )$ is
the dimension of $T_{\lambda }$, $\lambda \in \hat{G}$.

Let now $L^2(\hat{G})$ be the space of all matrix valued
functions on $\hat{G}$ such that: (i) $\phi(\lambda ) \in {\rm
Mat}_{n(\lambda )}({\bf C})$ for every $\lambda \in \hat{G}$,
(ii) $\sum_{\lambda \in \hat{\Gamma }}n(\lambda){\rm tr
}(\phi(\lambda)^*,\phi(\lambda )) < \infty$.  $L^2(\hat{G})$ is
a Hilbert space with the scalar product $$ (\phi_1,\phi_2) =
\sum_{\lambda \in \hat{\Gamma}}n(\lambda ){\rm
tr}(\phi_1(\lambda )\phi_2(\lambda )^*).  $$ It can be shown
that the Fourier transform of $f \in L^1(G , dg)$ prolongs to
the isometric mapping of the space $L^2(G, dg)$ onto the space
$L^2(\hat{G})$ (Ref. 23, p.194).

Basing on these results we define the {\it dual groupoid\/}
$\hat{\Gamma } = E \times \hat{G}$ of the groupoid $\Gamma =E
\times G$. Accordingly, we have the ``dual algebra'' $\hat{\calA
} = \{\hat{\Gamma} \rightarrow {\rm End}\calH_{\calA}\}$ to the
algebra $\calA = \{G \rightarrow {\bf C} \}$. Let $a\in \calA
,\; a=(a_p)_{p\in E}$. On the strength of the Fourier transform
properties to every $a_p: \Gamma_p \rightarrow {\bf C}$ there
corresponds the function $$
\hat{a_p}: \hat{\Gamma}_p \rightarrow \bigsqcup_{\lambda \in
\hat{\Gamma }_p}{\rm Mat}_{n(\lambda )}({\bf C}),
$$ where $\bigsqcup $ denotes disjoint sum, and $\hat{a} =
(\hat{a}_p)_{p\in E}$. Since we have the representation $\pi_p $
of the algebra \calA \ in the Hilbert space $L^2(\Gamma_p), \,
p\in E$, $\pi_p\!: \calA \rightarrow {\rm End }L^2(\Gamma_p)$,
we obtain the Fourier transform $$ L^2(\Gamma_p) \stackrel{\cal
F}{\rightarrow } L^2(\hat{\Gamma }_p) $$ where
$L^2(\hat{\Gamma}_p)$ is the Hilbert space with the scalar
product $$ (\phi_1, \phi_2) = \sum_{\lambda \in \hat{\Gamma
}}n(\lambda ){\rm tr}(\phi_1(\lambda ), \phi_2 (\lambda )^*).
$$

The algebra $\hat{\calA }$ can be regarded as leading to the
momentum representation of our model. In this representation a
projection corresponds to the momentum operator and a derivation
to the position operator. If $G$ is a compact group one can use
the Peter-Weyl theorem\cite{ref26} to develop the theory of dual
groupoids.

\section{Corresppondence with Stan\-dard Theories}
Since our model is a unification of general relativity with
quantum mechanics, we should check whether it leads to these
theories as its suitable ``limiting cases''. We do this in the
present section. Additionally, we discuss the transition from
the quantum sector of our model to classical mechanics.  It
turns out that all these transitions beautifully fit into the
structure of the model. To see the importance of such
discussions we send the reader to an interesting paper by Joy
Christian
\cite{ref27}.

\subsection{Correspondence: The Model -- General Relativity}
The transition from our model to the standard theory of general
relativity is done essentially by restricting the algebra $
{\cal A}={\cal A}_c\times {\cal A}_{cons t}$ to the subalgebra
${\cal A}_{proj}\subset {\cal A}_{const}$. If we assume (as in
Section V.A) that ${\cal A}_{ const}$ consists only of bounded
functions then ${\cal A}_{ proj}=$$\pi^{*}_M(C_b(M))$ where
$C_b(M) \subset C^{\infty}(M)$ denotes all smooth, bounded,
complex valued functions on $M$.  It is obvious that from $
C^{\infty}(M)$ one can reconstruct the full geometry of
space-time $ M$. In this way, one recovers the standard theory
of general relativity (see Ref. 3). We shall show that this can
also be achieved if we use $ C_b(M)$ instead of $C^{\infty}(M)$.

Let $M$ be a non empty set and $C$ a family of functions on $
M$. We denote by sc$C$ the set of functions on $ M$ of the form
$\omega\circ (\alpha_1,\ldots ,\alpha_n)\in C$ where $\omega$ is
a smooth function on ${\bf R}^n$ and $\alpha_ 1,\ldots
,\alpha_n\in C,\,$$n=1,2,\ldots$ A family $C$ is said to be {\em
closed with respect to the smooth functions\/} on $ {\bf R}^n$
if $C={\rm s}{\rm c}C$.

Let $\tau_C$ be the weakest topology on $ M$ in which all
functions belonging to $C$ are continuous, and let $ A$ be a
subset of $M$. By $C_M$ we denote the set of functions
$g:A\rightarrow {\bf R}$ such that for each point $ p\in^{}A$
there exists an open neighborhood $U$ of $ p$ and a function
$f\in C$ such that $f| U\cap A=g|U\cap A$. A set $C$ is said to
be {\em closed with respect to localization\/} if $ C_M=C$. If
$C=({\rm s}{\rm c}C)_M$ then the pair $( M,C)$ is called a {\em
differential space\/} and $C$ a {\em differential structure\/}
on $M$.

Let $(M,C)$ be a differential space. Its differential structure
is said to be {\em finitely generated\/} by a set $
C_0=\{\alpha_1,\ldots ,\alpha_n)$ if $({\rm s} {\rm c}C_0)_M=C$.

It can be easily seen that the differential structure $
C^{\infty}({\bf R})$ of the differential space $ $$ ($${\bf
R}^{},C^{\infty}({\bf R}))$ (which is, of course, a differential
manifold) is generated by the identity on {\bf R}, $i d_{{\bf
R}}$, which is an unbounded function, but it can also be
generated by a bounded function, for instance by $x\mapsto {\rm
a} {\rm r}{\rm c}{\rm t}{\rm g}x,x\in {\bf R}$.  Indeed, each of
these two functions can be expressed by the other one by the
composition with a function of the class $C^{\infty}$. The same
is true for the differential space $({\bf R}^n,C^{\infty} ({\bf
R}^n))$. Since any smooth manifold is locally a Euclidean space,
its differential structure is generated by coordinates of a
local map. Their composition with the function $ {\rm a}{\rm
r}{\rm c}{\rm t}{\rm g}x$ gives us the set of bounded functions
that generates this differential structure. Therefore, without
any loss of generality, we can assume that $ {\cal A}_{proj}$
consists of bounded functions.

It seems rather unusual that the transition from our model to
the ordinary space-time geometry has the character of a
restriction of a certain algebra to its subalgebra. In physics,
in analogous situations, we are usually confronted with a
``smooth transition to a limit''. We might, however, regard the
restriction of our algebra as a ``phenomenological description''
of a more subtle process.  Indeed, in the case when $G$ is a
noncompact group, on the strength of Proposition II.1, every
function $f\in {\cal A}_{proj}$ can be regarded as a (pointwise)
limit of a sequence of functions belonging to the algebra ${\cal
A}_c$. In this sense, also in our model, the emergence of
space-time can be thought of as a ``limiting process''.

\subsection{Correspondence: The Model -- Quantum Mechanics}
The quantum sector of our model, in spite of many similarities,
in some respects differs from the standard quantum mechanics.
First of all, it is strongly coupled to gravity. Formally, this
has been achieved by the fact that both gravity and quantum
effects are modelled by the same algebra on the groupoid $
\Gamma $ (kinematic aspect), and the fact that derivations
determining the dynamics of quantum operators (Eq.
(\ref{dyneq})) are postulated to be solutions of Einstein
equation (\ref{Einstein}) (dynamical aspect). As the consequence
of these properties, the phase space of our ``generalized
quantum mechanics'' is the Hilbert space $ {\cal
H}_q=L^2(\Gamma_q)$ (or ${\cal H}=\bigoplus_{ q\in
E}L^2(\Gamma_{q)})$, and since ${\cal H}_ q$, for all $q\in E,$
is isomorphic with $L^2(G)$, in fact, we have a ``quantum
mechanics on the group $G$'' rather than on $ {\bf R}^n$, as it
is the case for the position representation of the ordinary
quantum mechanics. Our model is strongly global, therefore, we
could expect that a ``local version'' of the model would lead to
the standard quantum mechanics. This is indeed the case. The
local version of the groupoid is the algebroid. We shall show
that the ``algebroid version'' of our model reproduces the usual
quantum mechanics.

The {\em Lie algebroid\/} of a Lie groupoid $
\Gamma $ is the vector bundle
\[A(\Gamma )=\bigcup_{\gamma\in \Gamma^0}T_{\gamma}
(\Gamma^p),\] where $\gamma =(p,e)$, with the bundle projection
$ \phi :A(\Gamma )\rightarrow \Gamma^0$ such that $\phi
(T_{\gamma}(\Gamma^p))=\gamma$. The fiber of this vector bundle
over $\gamma =(p,e)\in \Gamma^0$ is, in fact, the tangent space
to $p$ along the space $
\Gamma^p$. The smooth structure of $A(\Gamma )$ is defined in
the natural way (see Ref. 15, p. 58). It can be easily seen that
when $\Gamma $ is a Lie group (which is trivially a groupoid)
then $\Gamma^0=\{e\}$, and the Lie algebroid of $\Gamma $
becomes the Lie algebra of $\Gamma $. Let us apply this to our
case.

Let $U_q$ be a (starshaped) neighbourhood of the unit element $
e$ of the group $G$. The exponential map
\[\psi_q\equiv {\rm e}{\rm x}{\rm p}:\,T_
eG \rightarrow U_q\] induces the isomorphism of the Hilbert
spaces
\[\psi^{*}_q:L^2(U_q)\rightarrow L^2(T_eG).\]
Indeed, the equality $\langle\xi ,\zeta\rangle
=\langle\xi\circ\psi_q,\zeta\circ\psi_q\rangle ,\,\xi\,,\zeta\in
L^2(U_q)$, follows from the construction of the integral on a
manifold in the local map $\psi_q$.

Let us consider the mappings
\[{\cal A}\stackrel {\pi_q}{\longrightarrow}
{\rm E}{\rm n}{\rm d}({\cal H}_q)\stackrel
J{\longrightarrow}{\rm End}(\bar {{\cal H}}_ q)\] where $\bar
{{\cal H}}_q=L^2(T_e\Gamma^q)$, and $J$ is defined in the
following way. First, let us introduce the commutative diagram

\begin{center}

\unitlength=1.00mm
\special{em:linewidth 0.4pt}
\linethickness{0.4pt}
\begin{picture}(87.00,104.00)
\put(25.00,100.00){\vector(1,0){50.00}}
\put(25.00,50.00){\vector(1,0){50.00}}
\put(18.00,95.00){\vector(0,-1){40.00}}
\put(80.00,95.00){\vector(0,-1){40.00}}
\put(18.00,100.00){\makebox(0,0)[cc]{${\cal H}_q$}}
\put(80.00,100.00){\makebox(0,0)[cc]{${\cal H}_q$}}
\put(18.00,50.00){\makebox(0,0)[cc]{$\bar{\cal H}_q$}}
\put(80.00,50.00){\makebox(0,0)[cc]{$\bar{\cal H}_q$}}
\put(50.00,104.00){\makebox(0,0)[cc]{$\Omega $}}
\put(50.00,54.00){\makebox(0,0)[cc]{$\bar{\Omega }$}}
\put(10.00,75.00){\makebox(0,0)[cc]{$\psi ^*_q$}}
\put(87.00,75.00){\makebox(0,0)[cc]{$\psi ^*_q$}}
\end{picture}

\end{center}
\vspace{-4.5cm}

\noindent
where $\Omega\in {\rm E}{\rm n}{\rm d}({\cal H}_ q)$ and
$\bar{\Omega}\in {\rm End}(\bar {{\cal H}}_ q)$. We have
\[\bar{\Omega }(\bar{\xi })=\Omega\xi\circ
\psi_q\]
where $\bar{\xi }=\xi\circ\psi_q$, and the mapping $ J$ is
defined by $J(\Omega )=\bar{\Omega}$. We thus have a ``local (or
algebroid) version'' of the representation $
\pi_q$ of the algebra ${\cal A}$,
$\bar{\pi}_q:{\cal A}\rightarrow {\cal B} (\bar {{\cal H}}_q)$
which is given by
\[\bar{\pi}_q=J\circ\pi_q.\]

By using this representation we can write the ``local version''
of our dynamical equation (\ref{dyneq})
\[i\hbar\bar{\pi}_q(v(a))=[F_v,\bar{\pi}_
q(a)].\]

The Hilbert space ${\cal H}_q=L^2(T_e\Gamma_q )$ is now defined
on the space ${\bf R}^n${\bf .} To complete the transition to
the ordinary quantum mechanics let us consider the equivalence
relation
\[(p_1,g_1)\sim (p_2,g_2)\Leftrightarrow
\exists_{g\in G}p_2=p_1g.\]
The function $\varphi\in {\cal A}$ will be called $ G$-{\it
invariant} if
\[(p_1,g_1)\sim (p_2,g_2)\Rightarrow\varphi 
(p_1,g_1)=\varphi (p_2,g_2).\] The set of all such functions
forms a subalgebra ${\cal A}_{inv}$ of $ {\cal A}$.
Accordingly, a derivation $v\in V_1$ will be called $G$-{\em
invariant\/} if there exist $
\varphi\in {\cal A}$ such that 
$v(\varphi )\in {\cal A}_{inv}$. The set of all $ G$-invariant
derivations of ${\cal A}$ will be denoted by $V_{inv}$.

Let $\frac {\partial}{\partial x_i}$  be a basis in $ {\bf
R}^n\;(=T_eG)$. Then our dynamical equation assumes the form
\[i\hbar\widehat{\frac {\partial a}{\partial 
x^i}}=[F_{\widehat{\frac {\partial }{\partial x_i}}},\bar{\pi
}(a)]\] where $\widehat{\frac {\partial a}{\partial
x^i}}=\bar{\pi}_q(\frac {\partial}{\partial x^i}(a))$. If we put
$F_{\widehat{\frac {\partial }{\partial x^0}}}=H$ where $H$ is
bounded and positive then we can identify it with the
Hamiltonian of the system, and we finally obtain
\[i\hbar\widehat{\frac {\partial a}{\partial 
x^0}}=[H,\hat {a}],\] The remaining components give us
\[i\hbar\widehat{\frac {\partial a}{\partial 
x^i}}=[F_{\widehat{\frac {\partial }{\partial x^i}}},\hat
{a}],\] with $i=1,2, \ldots ,n$, and $\hat {a}=\bar{\pi } (a)$.
The last two equations are the well known quantum mechanical
equations for the evolution of energy and momentum,
respectively.

\subsection{Transition: Quantum Mechanics -- Classical Mechanics}
The transition form quantum mechanics to classical mechanics is
a standard problem which can be discussed (and is widely
discussed) beyond the framework of our model. It turns out,
however, that our model places this transition in a transparent
conceptual setting and within a natural mathematical structure.
The idea is simple: if in our model we go from the groupoid
$\Gamma $ to its tangent groupoid ${\cal G}_{\Gamma }$ then our
formulation of quantum mechanics goes to the usual classical
mechanics.  Although this procedure is known and was applied to
the standard formulation of quantum mechanics (see, for example,
Ref. 15, pp. 78-84), we repeat it briefly for the sake of
completeness of this presentation.

Let us define ${\Gamma }_{\epsilon}:=({\Gamma }\times {\Gamma
})\times\epsilon$ where $\epsilon\in {\bf R}\setminus \{0\}$,
and ${\Gamma }_0:=T{\Gamma }\times \{0\}$ where $T{\Gamma }$ is
the tangent bundle. The {\em tangent groupoid\/} is defined to
be
\[{\cal G}_{\Gamma }:=\bigcup_{\epsilon}{\Gamma }_{\epsilon}
\cup {\Gamma }_0\]
with the usual topologies on ${\Gamma }_{\epsilon}$ and $
{\Gamma }_0$, and the condition that if $\epsilon\rightarrow 0$
then ${\Gamma }_{\epsilon}
\rightarrow {\Gamma }_0$ in the following sense: for any sequence $
(\gamma_n,\eta_n,\epsilon_n)$ such that $\gamma_n,\,$$\eta_n\in
{\Gamma }$, if $\lim_{ n\rightarrow\infty}\eta_n=\eta
=\lim\gamma_ n$, $\lim_{n\rightarrow\infty}\epsilon_n =0$, and
$\lim_{n\rightarrow\infty}\frac {\gamma_
n-\eta_n}{\epsilon_n}=X\in T{\Gamma }$, then
\[(\gamma_n,\eta_m,\epsilon_n)\rightarrow 
(\eta ,X,0)\in T{\Gamma }\times \{0\}.\] It can be shown that
the $C^{*}$-algebra of the groupoid $ {\Gamma }_{\epsilon}$ is
the algebra ${\cal K}(L^2({\Gamma }))$ of compact operators on
the separable Hilbert space $ L^2({\Gamma }),$ and the
$C^{*}$-algebra of the groupoid ${\Gamma }_0$ is $
C_0(T^{*}{\Gamma })$ (Ref. 15, p. 79). The tangent groupoid
structure gives us the following deformation of groupoids
${\Gamma }_{\epsilon}\stackrel {\epsilon\rightarrow
0}{\longrightarrow}{\Gamma }_0$ and, consequently, the following
deformation of their $ C^{*}$-algebras
\[{\cal K}(L^2({\Gamma }))\stackrel {\epsilon\rightarrow 
0}{\longrightarrow}C_0(T^{*}{\Gamma }).\]

Let us notice that we have the inclusion
\[\beta :C_0(T^{*}{\Gamma })\hookrightarrow C_0(
T^{*}M).\] Indeed, the natural projection $pr_M:{\Gamma
}\rightarrow M$ leads to the mapping
\[pr_M^{*}:T^{*}M \rightarrow T^{*}{\Gamma }\]
such that to each function $f\in C_0(T^{ *}{\Gamma })$ there
corresponds the function $f\circ\pi_M^{*}\in C_0(T^{*}M)$.

If we now choose the subalgebra $\tilde { {\cal A}}$ of our
algebra ${\cal A}$ such that $\pi (\tilde {{\cal A}})={\cal
K}(L^2({\Gamma }) )$, where $\pi$ is the representation
(\ref{reprpi}) of $ {\cal A}$, we obtain the deformation
\[\pi (\tilde {{\cal A}})={\cal K}(L^2({\Gamma }
))\stackrel {\epsilon\rightarrow 0}{\longrightarrow}
C_0(T^{*}{\Gamma })\stackrel {\beta}{\hookrightarrow}
C_0(T^{*}M).\] If we notice that ${\cal K}(L^2({\Gamma }))$ is
the algebra of observables of the quantum sector of our model,
and $C_0(T^{*}M)$ the algebra of observables of classical
mechanics, it is enough to identify the parameter $
\epsilon$ with the Planck constant $\hbar$ to obtain the
transition from the quantum sector of our model to classical
mechanics. Let us also notice that
\[C_0(T^{*}M)\setminus\beta (C_0(T^{*}{\Gamma })
)\neq\emptyset\] implies that there exist classical observables
that do not come from compact quantum operators; for instance,
position and momentum in quantum mechanics are non compact
operators.

\section{Noncommutative Fock Space}
The model constructed in the previous sections gives us a
conceptually transparent unification of general relativity and
quantum mechanics.  However, its main drawback is that it is
lacking the quantum field theoretical aspect. To at least
partially improve this situation and to indicate the line of the
future development, in the present Section, we construct the
Fock space for our model.

As it is well known, Fock space is a phase space for a quantum
system of many identical, noninteracting particles which in the
following will also be called quanta. Let ${\cal H}$ be a space
of states of a single quantum.  In our model $ {\cal H}$ is
$L^2(\Gamma_p)$ or, if needed, $L^2(\Gamma)=\bigoplus_{ p\in
E}L^2(\Gamma_p),\,p\in E$.  We assume that {\bf C} is the space
of vacuum states, i.  e.,  the space of ``zero quantum states''.
Let us define the direct sum of the following spaces (as linear
spaces) $F={\bf C}\oplus {\cal H}\oplus ({\cal H}\otimes {\cal
H})\oplus ({\cal H}\otimes {\cal H}\otimes {\cal
H})\oplus\cdots$ (${\bf C}$ and ${\cal H}$ are regarded here as
linear spaces). This space would correspond to a superposition
of states having different numbers of quanta (no quantum, one
quantum, two quanta...),  but it is too large; we must correct
it by taking into account that quanta are indistinguishable.  To
this end we define the {\em symmetrization operator\/} $\sigma
={\cal H} \otimes \cdots
\otimes {\calH}\rightarrow {\cal H}\otimes \cdots \otimes {\cal
H}$ by
\[\sigma (\xi_1\otimes\cdots\otimes
\xi_n)=\frac 1{n!}\sum_{s\in\Pi_
n}\xi_{s(1)}\otimes^{}\cdots\otimes
\xi_{s(n)}\]
where $\Pi_n$ is the set of $n$ elementary permutations; and the
{\em antisymmetrization operator} $\tau ={\cal H}\otimes\cdots
\otimes {\cal H}\rightarrow {\cal H}\otimes
\cdots \otimes {\cal H}$ by
\[\tau (\xi_1\otimes\cdots\otimes
\xi_n)=\frac 1{n!}\sum_{s\in\Pi_
n}{\rm s}{\rm i}{\rm g}{\rm n}
(\xi_{s(1)}\otimes^{}\cdots\otimes
\xi_{s(n)}).\]
Of course, $\sigma\circ\sigma =\sigma $, and $\tau\circ\tau
=\tau $.  An element $w\in {\cal H}\otimes
\cdots\otimes {\cal H}$ is {\em symmetric\/} if $
\sigma (w)=w$, and {\em antisymmetric\/} if 
$\tau (w)=w.$ Let us denote by $ F_{\sigma}$ the subspace of the
vector space $ F$ all components of which are symmetric, and by
$ F_{\tau}$ the subspace of $F$ all components of which are
antisymmetric. $ F_{\sigma}$ is called the {\em symmetric Fock
space,\/} and it describes bosonic quantum states. $ F_{\tau}$
is called the {\em antisymmetric Fock space}, and it describes
fermionic quantum states.

So far it was simply a repetition of the well known construction
of the Fock space for the case when ${\cal H} =L^2(\Gamma_p)$ or
${\cal H}=L^2(\Gamma )$. The rest is also straightforward, but
it requires a certain care. We must check whether the analogous
construction can be carried out for the algebra \calA \ (if
necessary regarded as a vector space).

First, let us consider the tensor product $ {\cal A}\otimes
{\cal A}$. Addition in this algebra is defined as usual:
homogeneous elements are added in the ordinary way whereas
nonhomogeneous elements form the internal direct sum. Involution
is defined in the natural way: $ (a_1\otimes
a_2)^{*}=a_2^{*}\otimes a_1^{*}$. If ${\cal A}$ is regarded as a
vector space, one can define the symmetrization and
antisymmetrization operations as above.

For any derivation $v\in {\rm D} {\rm e}{\rm r}{\cal A}$ we
define the derivation of the algebra $ {\cal A}^p$, $\bar
{v}:{\cal A}^p\rightarrow {\cal A}^p$ by
\[\bar {v}(a_1\otimes\cdots\otimes 
a_p)=\sum_{i=1}^pa_1\otimes\cdots
\otimes a_{i-1}\otimes v(a^{}_
1)\otimes a_{i+1}\otimes\cdots
\otimes a_p.\]

Now we can construct the {\em Fock algebra\/} in the following
way
\[F({\cal A})={\bf C}\oplus {\cal A}
\oplus {\cal A}^2\oplus\cdots\oplus 
{\cal A}^p\oplus\cdots\]

In the following we shall consider the ``extended module'' $
\bar {V}$ of derivations of the Fock algebra. We thus have the
differential algebra $ (F({\cal A}),\bar {V})$ corresponding to
the Fock space; we shall call it the {\em Fock differential
algebra}. Basing on this differential algebra we shall construct
the geometry of the Fock space.

For any tensor $T:V\times\cdots\times V\rightarrow V$ we define
its extension $\bar {T}:\bar {V}\times\cdots
\times\bar {V}\rightarrow\bar {V}$ by
\[\bar {T}(v_1,\ldots ,v_k)=\overline {
T(v_1,\ldots ,v_k)}.\]

From the above we immediately have the following lemma.
\begin{lemma}
$\bar {T}=0$ if and only if $ T=0$. $\Box$
\end{lemma}

Since the Einstein equation of our model has the form $ {\bf
G}=0$, where {\bf G} is the generalized Einstein tensor, the
above lemma implies that the same form of the Einstein equation
is valid on the Fock space.

It is straightforward that the  representation of the algebra $
{\cal A}^p$ in the Hilbert space ${\cal H}^p=\bigotimes_{
i=1}^p{\cal H}_i$, $\pi_q:{\cal A}^p\rightarrow {\rm E}{\rm
n}{\rm d}{\cal H}^p$ should be defined by
\[\pi_q(a_1\otimes\cdots\otimes 
a_p):=\pi_q(a_1)\otimes\cdots\otimes
\pi_q(a_p)\]
where $\pi_q(a_i),\,i=1,\ldots ,p$, is the representation of the
algebra $ {\cal A}$ in the Hilbert space ${\cal H}$ given by
(\ref{reprpiq}). For instance, on a simple element we have
\[\pi_q(a_1\otimes\cdots\otimes 
a_p)(\xi_1\otimes\cdots\otimes
\xi_p)=\pi_q(a_1)(\xi_1)\otimes
\cdots\otimes\pi_q(a_p)(\xi_p)
=\]
\[=(a_1\otimes\cdots\otimes a_
p)*(\xi_1\otimes\cdots\otimes\xi_ p).\]

Now, it is easy to write dynamical equation (\ref{dyneq}); for
instance, in gradation 2 we have
\[i\hbar\pi_q(\bar {v}(a_1\otimes^{}
a_2))(\xi_1\otimes\xi_2)=[F_v\otimes F_v,\pi_q(a_1\otimes
a_2)](\xi_ 1\otimes\xi_2)\] with $\bar {v}\in {\rm k}{\rm e}
{\rm r}\bar {{\bf G}}$ and all other symbols self-evident.

An element $\xi_1\otimes\cdots
\otimes\xi_p$ is said to be $G${\em -invariant\/} if all
its components are $G$-invariant (i. e., if they are constant on
the equivalence classes of the action of the group $G$). Let
$a=(a_ 1\otimes\cdots\otimes a_p)\in {\cal A}_{G}\otimes\cdots
\otimes {\cal A}_{G}$, and let $
\tilde{\xi}$$_1\otimes\cdots\otimes
\tilde{\xi}_p$ be $G$ invariant. In such a case $
\tilde{\xi}$$_1\otimes\cdots\otimes
\tilde{\xi}_p$ can be regarded as an element of the 
Fock space $F(L^2(\Gamma_q)$, and the equation
$$ i\hbar\pi_q(\bar {v}(a_1\otimes
\cdots\otimes a_p))(\tilde{
\xi}_1\otimes\cdots\otimes\tilde{
\xi}_p)= $$
$$ [F_v\otimes\cdots\otimes F_v, \pi_q(a_1 \otimes \cdots
\otimes a_p)](\tilde{\xi}_1\otimes\cdots
\otimes\tilde{\xi}_p)
$$ 
space.

The standard operators on the above noncommutative Fock space
can be defined in close analogy to the usual case. First, we
define the {\em number operator} $N:F({\cal H})\rightarrow
F({\cal H} )$, by
\[\xi =(\xi_0,\xi_1,\xi_2,\ldots 
)\mapsto N(\xi )=(0,1,\xi_1,2\xi_ 2,3\xi_3,\ldots )\] where
$\xi_i\in\bigotimes_{i=1}^ p{\cal H}^p.$ It can be easily seen
that there exists $(a_0,a_1,a_2,\ldots )\in F({\cal A} )$, where
$a^p\in {\cal A}$, such that $
\pi_q(a_i)=\xi_i$, and we have the 
corresponding operator in the Fock algebra $\tilde {N}:F({\cal
A})\rightarrow F({\cal A})$ defined by
\[\tilde{N}(a_0,a_1,a_2,\ldots 
)=(0,1a,2a,3a,\ldots ).\]

If $\xi$ is in the symmetric Fock space $ F_{\sigma}$, so is
$N(\xi )$; the same refers to the antisymmetric Fock space
$F_{\tau}$. Therefore, in fact we have two linear operators:
$N_{\sigma}$ in $F_{
\sigma}$, and $N_{\tau}$ in $F_{
\tau}$. For instance, for $\xi
\in F_{\sigma}$, one has 
$N(\xi )=3$ if and only if $\xi =(0,0,0,\xi_3)$ which means that
the system is in the state corresponding to 3 bosons.

Let us fix $\eta\in {\cal H}=L^ 2(G_q)$, and let us define the
{\em creation operator} $C_{\eta}:F({\cal H})\rightarrow F({\cal
H})$ by
\[C_{\eta}(\xi )=(0,\sqrt {1}\xi_
0\eta ,\sqrt {2}\eta\otimes\xi_ 1,\sqrt
{3}\eta\otimes\xi_2,\ldots ),\] and correspondingly for the Fock
algebra $\tilde {C}_a:F({\cal A})\rightarrow F({\cal A})$ by
\[\tilde {C}_b(a_0,a_1,a_2,a_3
,\ldots )=(0,\sqrt {1}a_0b,\sqrt { 2}b\otimes a_1,\sqrt
{3}b\otimes a_2,\ldots )\] where $a$ is a fixed element of $
{\cal A}$. It can be checked that
\[N\circ C_{\eta}=C_{\eta}\circ 
(N+I_F),\] where $I_F$ is the identity on $ F$; and analogously
\[\tilde {N}\circ\tilde {C}_b=
\tilde {C}_b\circ (\tilde {N}+
I_{F({\cal A})}).\] The above formula says that the action of
the creation operator increases the number of quanta by one.

Let $\phi$ be an element of the dual $ {\cal H}^{*}$ to the
Hilbert space $ {\cal H}.$ The {\em annihilation $operator$ }
$A_{\phi}:F({\cal H})\rightarrow F({\cal A})$ is defined by
\[A_{\phi}(0,0,\ldots ,\zeta\otimes
\zeta_1\otimes\cdots\otimes\zeta_{
n-1},0,0)=(0,0,\ldots\sqrt n\phi (\zeta
)\otimes\cdots\otimes\zeta_{ n-1},0,0,\ldots ),\] and then one
extends (by linearity) the operator $ A_{\phi}$ to the whole
space $F({\cal H})$. Analogously one defines the corresponding
annihilation operator for the Fock algebra $\tilde
{A}_{\varphi}:F({\cal A} )\rightarrow F({\cal A})$ by
\[\tilde {A}_{\varphi}(0,0,\ldots 
,b\otimes a_1\otimes\cdots\otimes a_{n-1},0,0)=(0,0,\ldots
,\sqrt n\phi (b)a_1\otimes\cdots\otimes a_{n-1},0,0\ldots )\]
for a fixed element $\varphi\in {\cal A}^{*}$. One also has
\[(\tilde {N}+I_{F({\cal A})})
\circ\tilde {{\cal A}}_{\varphi}
=\tilde {{\cal A}}_{\varphi}\circ
\tilde {N}\]
(and similarly for the space $ F({\cal H})$) which says that the
action of the annihilation operator diminishes the number of
quanta by one.

It can be easily checked that the following (anti)commutation
relations (analogous to those known from the standard field
theory) are valid:

For the bosonic sector
\[[\tilde {C}_b,\tilde {C}_{b'}
]=0, \;\; [\tilde {A}_{\varphi},\tilde { A}_{\varphi'}]=0,\;\;
[\tilde {A}_{\varphi},\tilde { C}_b]=[\varphi (b)]I_{F_{\sigma}
({\cal A})},\]

and for the fermionic sector
\[[\tilde {C}_b,\tilde {C}_{b'}
]_{+}=0,\;\; [\tilde {A}_{\varphi},\tilde {
A}_{\varphi'}]_{+}=0,\;\; [\tilde {A}_{\varphi},\tilde {
C}_b]=[\varphi (b)]I_{F_{\tau} ({\cal A})}.\]

\section{Concluding Remark} 
The model developed in the present paper creates a transparent
conceptual framework for a unification of general relativity and
quantum mechanics. It ``predicts'' nonlocal phenomena present in
both these theories, such as the horizon paradox and the EPR
experiment; it explains the state vector reduction, and
naturally unifies probability and dynamics (both generalized
probability and dynamics are incorporated into von Neumann
algebras, see Ref.  2). It also has the correct correspondence
with general relativity, quantum mechanics, and classical
mechanics.  And the construction of the Fock space for this
model suggests that it can be enlarged to include the field
theoretical aspect. Two elements, on which the model is based,
are responsible for these results. First of them is obviously
the idea of noncommutative geometry. It has made the model
consequently nonlocal and enabled the generalization of many
geometrical concepts. The second element is the groupoid
structure. Owing to it the standard physical concept of symmetry
has been replaced by a generalized symmetry. The generalization
is at least twofold: first, groupoids can be thought of as
``groups with many identities''; second, an algebraic structure
underlying geometry based on a group is a set with a
distinguished base point, whereas the analogous structure
underlying a groupoid geometry is a directed graph
\cite{Bangor}. This opens new possibilities for a physical
theory. Some of them were explored in the present paper.

\end{document}